\documentclass[12pt]{article}
\pdfoutput=1
\usepackage{amsmath}
\usepackage{amssymb} 
\usepackage{bbm} 
\usepackage[small]{caption2} 
\usepackage{fleqn} 
\usepackage{graphicx} 
\usepackage{mathrsfs}   
\usepackage[small,loose]{subfigure}  
\usepackage{cite} 
\usepackage{color}
\usepackage{xspace}
\usepackage{graphicx}
\usepackage{color}
\usepackage{ulem}

\newcommand{\eVdist}{\kern-0.06em}

\newcommand{\gev}{\:\text{Ge\eVdist V}}
\newcommand{\tev}{\:\text{Te\eVdist V}}

\newcommand{\Z}[1]{\ensuremath{\mathbbm{Z}_{#1}}} 

\newcommand{\hu}{\ensuremath{H_{u}}}
\newcommand{\hd}{\ensuremath{H_{d}}}

\newcommand{\singlet}{\ensuremath{S}}

\newcommand{\be}{\begin{equation}}
\newcommand{\ee}{\end{equation}}
\newcommand{\bea}{\begin{eqnarray}}
\newcommand{\eea}{\end{eqnarray}}

\newcommand{\SARAH}{{\tt SARAH}\xspace}
\newcommand{\SPheno}{{\tt SPheno}\xspace}
\newcommand{\MO}{{\tt MicrOmegas}\xspace}
\newcommand{\HB}{{\tt HiggsBounds}\xspace}
\newcommand{\SSP}{{\tt SSP}\xspace}
\newcommand{\CalcHep}{{\tt CalcHep}\xspace}

\allowdisplaybreaks[1]

\begin{document}

\begin{titlepage}

\vspace*{-3.0cm}
\begin{flushright}
OUTP-12-06P\\
Bonn-TH-2012-04
\end{flushright}

\begin{center}
{\Large\bf
  The generalised NMSSM at one loop:\\ fine tuning and phenomenology
}

\vspace{1cm}

\textbf{
Graham G.~Ross$^a$,
Kai Schmidt-Hoberg$^a$,
Florian Staub$^b$
}
\\[5mm]
\textit{$^a$\small
Rudolf Peierls Centre for Theoretical Physics, University of Oxford,\\
1 Keble Road, Oxford OX1 3NP, UK
}
\\[5mm]
\textit{$^b$\small
Bethe Center for Theoretical Physics \& Physikalisches Institut der 
Universit\"at Bonn, \\
Nu{\ss}allee 12, 53115 Bonn, Germany
}
\end{center}

\vspace{1cm}

\begin{abstract}
We determine the degree of fine tuning needed in a generalised version of the NMSSM that follows from an underlying $\Z{4}$ or $\Z{8}$ $R$ symmetry.  We find that it is significantly less than is found in the MSSM or NMSSM and extends the range of Higgs mass that have acceptable fine tuning up to Higgs masses of $m_h \sim 130 \gev$. 
For universal boundary conditions analogous to the CMSSM the phenomenology is rather MSSM like with the singlet states typically rather heavy.
For more general boundary conditions the singlet states can be light, leading to interesting signatures at the LHC and direct detection experiments. 
\end{abstract}

\end{titlepage}

\section{Introduction}

The recent results from the ATLAS and CMS collaborations provide an indication, albeit inconclusive, that the Higgs may lie in the range 124-126 GeV \cite{ATLAS:2012ae,Chatrchyan:2012tx,ATLAS:2012ad,ATLAS:2012ac, Chatrchyan:2012tw, Chatrchyan:2012dg, Chatrchyan:2012ty, Chatrchyan:2012sn}. A mass in this range has significant implications for supersymmetric extensions of the Standard Model (SM) capable of ameliorating the (little) hierarchy problem.  In particular in the Constrained Minimal Supersymmetric Standard Model (CMSSM) the fine tuning needed to achieve this mass is large, requiring a cancellation between uncorrelated parameters of order 1 part in 300. In the more general context of the MSSM one still requires 1\% fine tuning even for an extremely low messenger scale of 10 TeV~\cite{Hall:2011aa}\footnote{Note that this definition of fine tuning differs from ours in the choice of the measure and the fact that the parameters are taken to be low-scale parameters.}. 

To accommodate a heavier Higgs while avoiding very large fine tuning requires new structure.  For the case this results from new physics at a scale higher than the supersymmetry breaking scale one can perform a model independent analysis through the inclusion of higher dimension operators \cite{Dine:2007xi, Cassel:2009ps}. This analysis showed that the fine tuning is extremely sensitive to an operator that is most simply generated through the integration out of a massive singlet chiral superfield that couples to the two Higgs supermultiplets of the MSSM. The effect of this operator is to allow a Higgs mass as high as $130 \gev$  without increasing the low fine tuning found in the MSSM. This applies even if the mass of the singlet states is as high as $3 \tev$. 

This structure is {\it not} the one found in the usual Next-to-Minimal-Supersymmetric-Standard-Model (NMSSM, see e.g.~\cite{Ellwanger:2006rm} for a review) which assumes there is an underlying $\Z{3}$ symmetry that forbids the singlet mass term and an explicit $\mu$ term. The reason this structure has been favoured is that it explains why the singlet should be light and, after supersymmetry breaking, the singlet superfield scalar component acquires a vacuum expectation value (vev) that generates a $\mu$ term of order the supersymmetry breaking scale. Fine-tuning studies of the NMSSM have been performed in e.g.~\cite{BasteroGil:2000bw,Dermisek:2005gg,Dermisek:2006py,Dermisek:2007yt,Ellwanger:2011mu}.  However recently it has been realised that if, instead of the $\Z{3}$ symmetry, the NMSSM has a discrete $R$ symmetry, $\Z{4}^R$ or $\Z{8}^R$, then, after supersymmetry breaking, both the singlet mass and the $\mu$ term are generated but both are constrained to be of order the supersymmetry breaking mass \cite{Lee:2010gv, Lee:2011dya}. Moreover the $R$ symmetry has the advantage that it forbids the dangerous dimension 5 proton decay operators\footnote{These operators, allowed in the MSSM and the usual NMSSM, must be suppressed by a mass of order $10^{7}$ times the Planck scale!}  and does not have the domain wall problem 
\cite{Abel:1995if} that is associated with the $\Z{3}$ symmetry of the normal NMSSM\footnote{Avoiding unacceptable domain walls requires an $R$ symmetry even for the usual NMSSM \cite{Abel:1996cr, Panagiotakopoulos:1998yw}}. To distinguish the discrete $R$ symmetric NMSSM from the usual NMSSM we will denote it by the GNMSSM~\cite{Ross:2011xv}.

In this paper we study the fine tuning in the GNMSSM in detail, without requiring the mass of the new singlet states be larger than the mass of the MSSM SUSY states. 
Also, in contrast to~\cite{Ross:2011xv}, we take into account the complete superpartner mass spectrum at one-loop
as well as all one-loop contributions and the dominant two-loop contributions in the Higgs sector. 
There is a broad minimum of the fine tuning for a Higgs mass between the LEP bound and $125 \gev$. We discuss the phenomenology associated with these low-fine tuned points and the implications for dark matter abundance and direct dark matter searches.

\section{The GNMSSM } 
\subsection{The superpotential}
The most general extension of the MSSM by a gauge singlet chiral superfield consistent with the SM gauge symmetry has a superpotential of the form
\begin{eqnarray}
 \mathcal{W} &=& \mathcal{W}_\text{Yukawa}  + \frac{1}{3}\kappa S^3+
(\mu + \lambda S) H_u H_d + \xi S+ \frac{1}{2} \mu_s S^2  \label{gen}\\
&\equiv& \mathcal{W}_\text{NMSSM}+
\mu H_u H_d + \xi S+ \frac{1}{2} \mu_s S^2  \label{gen2} 
\end{eqnarray}
where $\mathcal{W}_\text{Yukawa}$ is the MSSM superpotential generating the SM Yukawa couplings and $ \mathcal{W}_\text{NMSSM}$ is the normal NMSSM with a $\Z{3}$ symmetry. Here and in what follows capital letters refer to superfields while small letters refer to the corresponding scalar component.
One of the dimensionful parameters can be eliminated by a shift in the vev $v_s$. 
We use this freedom to set the linear term in $S$ in the superpotential to zero, $\xi=0$. 
Such a superpotential can arise from an underlying $\Z{4}^R$ or $\Z{8}^R$ symmetry \cite{Lee:2010gv,Lee:2011dya}.
Before SUSY breaking the superpotential is of the NMSSM form. However after supersymmetry breaking in a hidden sector with gravity mediation soft superpotential terms are generated but with a scale of order the supersymmetry breaking scale in the visible sector characterised by the gravitino mass, $m_{3/2}$. With these the renormalisable terms of the superpotential take the form \cite{Lee:2011dya}
\begin{eqnarray}
 \mathcal{W}_{\Z{4}^{R}}
 & \sim &    \mathcal{W}_\text{NMSSM}+ m_{3/2}^2\, \singlet + m_{3/2}\, \singlet^{2} 
               + m_{3/2}\, \hu\, \hd \;,\\
   \mathcal{W}_{\Z{8}^{R}}
  & \sim & \mathcal{W}_\text{NMSSM}+ m_{3/2}^2\, \singlet 
\label{eq:WNMSSM1}
\end{eqnarray}
where the $\sim$ denotes that the dimensional terms are specified up to $\mathcal{O}(1)$ coefficients.
Clearly the $\Z{4}^R$ case is equivalent to the GNMSSM. After eliminating the linear term in $S$ the $\Z{8}^R$ case gives a constrained version of the GNMSSM with $\mu_{s}/\mu=2\kappa/\lambda$.

Note that the SUSY breaking also breaks the discrete $R$ symmetry but leaves the subgroup $\Z{2}^{R}$, corresponding to the usual matter parity, unbroken. As a result the lightest supersymmetric particle, the LSP, is stable and a candidate for dark matter.

\subsection{Supersymmetry breaking}

The general soft SUSY breaking  terms associated with the Higgs and singlet sectors are
\begin{align}
 V_\text{soft} 
   &=  m_s^2 |s|^2 + m_{h_u}^2 |h_u|^2+ m_{h_d}^2 |h_d|^2 \nonumber \\
   &+ \left(b\mu \, h_u h_d + \lambda A_\lambda s h_u h_d + \frac{1}{3}\kappa A_\kappa s^3 + \frac{1}{2} b_s s^2  + \xi_s s + h.c.\right) \;.
\label{soft}
\end{align}
Note that the shift in the vev $v_{s}$ that is used to eliminate the linear term in the superpotential does not imply that the corresponding soft term $\xi_s$ is zero as well.

These terms and the soft breaking terms associated with the squarks, sleptons and gauginos depend on the details of the supersymmetry breaking sector. Here we will first consider the simplest case, the CGNMSSM, chosen in analogy with the well-known CMSSM, with a universal scalar and gaugino mass and all other soft terms proportional to their corresponding superpotential couplings. In order to compare to the usual NMSSM case we will subsequently relax the universality of scalar masses to allow the Higgs masses to differ.
We furthermore allow for independent $A_\lambda$, $A_\kappa$ in the second case.

The independent supersymmetry breaking parameters of the CGNMSSM are  
$m_0$, $m_{1/2}$, $A_0$, $B_0$ and $\xi_s$ where $A_0$ and $B_0$ are the constants of proportionality associated with the trilinear and bilinear terms respectively. These parameters are defined at the unification scale, $M_{X}$, and must be evaluated at low scales using the renormalisation group running.

Taking into account the supersymmetric parameters as well, the CGNMSSM within this simple supersymmetry breaking scheme is specified by the following set of parameters
$\mu$, $\mu_s$, $\lambda$, $\kappa$, $m_0$, $m_{1/2}$, $A_0$, $B_0$ and $\xi_s$ (in the more general case we have $m_{h_d}^2$, $m_{h_u}^2$, $m_{s}^2$, $A_\lambda$, $A_\kappa$ 
in addition). Trading $B_0$, $\xi_s$ and $\mu$ for $v$, $\tan\beta$ and $v_s$ via the EWSB conditions,
there are eight (thirteen) parameters defining these models.

\section{Fine-Tuning and dark matter abundance: analysis methods}
\subsection{The fine tuning measure}

As introduced in \cite{Ellis:1986yg, Barbieri:1987fn}, a quantitative estimate of the the fine tuning  with respect to a set of independent parameters, $p$,   is given by
\begin{equation} 
\Delta \equiv \max {\text{Abs}}\big[\Delta _{p}\big],\qquad \Delta _{p}\equiv \frac{\partial \ln
  v^{2}}{\partial \ln p} = \frac{p}{v^2}\frac{\partial v^2}{\partial p} \;.
\end{equation}
where $v$ is the EW scale\footnote{$v^{2}=v_{u}^{2}+v_{d}^{2}$ where $v_{u,d}$ are the up and down sector Higgs vacuum expectation values. Here we work in conventions
in which $v \simeq 246 \gev$.}.
The quantity $\Delta^{-1}$ gives a measure of the accuracy to which independent parameters must be tuned to get the correct electroweak breaking scale. The parameters, $p$, correspond to the nine (fourteen) independent parameters discussed above plus the top Yukawa coupling\footnote{We use the modified definition for fine tuning~\cite{Ciafaloni:1996zh} for the top-Yukawa coupling, appropriate for measured parameters.} all defined at the unification scale and chosen to be of mass dimension 2 where appropriate, e.g.~$\mu^{2}$.

\subsection{SUSY particle spectrum constraints from the LHC}

A large part of the low fine tuned parameter space is not viable because of the constraints from the LHC.
If all squarks are roughly degenerate, as is the case in models with a universal squark mass at the high scale,
the bounds on the squarks and gluinos are very stringent.
The precise bounds depend on the details of the sparticle spectrum. Bounds for some cases can e.g.\ be found online at~\cite{WebAtlas,WebCMS}. \\
The strongest bound assumes a light neutralino and that all squarks and gluinos are degenerate, giving $m_{\tilde{s}}=m_{\tilde{g}} > 1400 \;\gev$,
while if one assumes that the squarks are twice as heavy as the gluino, the bound on the gluino mass changes to $m_{\tilde{g}} \simeq 700 \gev$.
If these assumptions are violated the bounds can be significantly weaker, e.g.\ a single stop below  $250 \gev$ is still allowed!

To allow for these uncertainties we take the more conservative superpartner bounds to be $m > 1200 \gev$ for the first two generation squarks and the gluino and for the chargino we apply the bound $m_{\tilde{\chi}^+} > 94 \gev$ \cite{Nakamura:2010zzi}. LEP also constrains the masses of sleptons to be above $\mathcal{O}(100) \gev$. However these limits are always fulfilled when we apply the cut on the squark masses because we assume a unification of all squark and slepton masses at the GUT scale.

\subsection{Implementation in \SARAH and \SPheno}
To perform an exhaustive study of the GNMSSM we have used the public Mathematica package \SARAH \cite{Staub:2008uz,Staub:2009bi,Staub:2010jh}. \SARAH contains  a model file for a general singlet extended MSSM (SMSSM) which is easily adaptable to the GNMSSM. Using this model file \SARAH  analytically calculates all mass matrices, vertices as well as the two-loop Renormalization Group Equations (RGEs) and one-loop corrections to self-energies and  tadpoles. The calculation of the loop corrections is performed in $\overline{\text{DR}}$ scheme and 't Hooft gauge. The results can be used to get the entire one-loop corrected mass spectrum based on the approach first used in \cite{Pierce:1996zz}. For a detailed discussion of the use of this method to calculate the one-loop mass spectrum in extensions of the MSSM we refer to \cite{Staub:2010ty} and \cite{O'Leary:2011yq}.

\SARAH also provides an interface to produce modules for \SPheno \cite{Porod:2003um,Porod:2011nf}; all derived, analytical expressions are  exported to Fortran code which can be compiled together with the public \SPheno version. This offers the possibility to automatically get a fully-fledged spectrum calculator which outputs the SUSY one-loop mass spectrum based on a two-loop evaluation of all parameters as well as the widths and branching ratios of all sparticles and Higgs fields. The resulting \SPheno version writes, for the parameter point under consideration, input files for \HB \cite{Bechtle:2011sb} which can be used to check all existing collider constraints in the Higgs sector. We have extended  \SPheno  by incorporating routines to calculate the fine tuning. As cross check we have implemented the equivalent functions in a MSSM version written by \SARAH and compared the results with the fine tuning calculation of {\tt SoftSUSY} \cite{Allanach:2001kg}. In addition, we have linked the known dominant two-loop corrections  in the Higgs sector involving the strong coupling, $\alpha_s$,  and the third generation Yukawa couplings \cite{Brignole:2001jy,Brignole:2002bz,Dedes:2002dy,Dedes:2003km} included in the public version of \SPheno.  

For a comparison of the fine tuning within the NMSSM we have used the \SPheno version presented in \cite{Staub:2010ty}, which includes also the known two-loop corrections of the NMSSM \cite{Degrassi:2009yq}, and implemented the routines for the calculation of the fine tuning. 

The dark matter relic density has been calculated with \MO \cite{Belanger:2006is,Belanger:2007zz,Belanger:2010pz}. To this end we used the option of \SARAH to write model files for \CalcHep \cite{Pukhov:2004ca} which can also be used with \MO. To interface \SPheno and \MO we used the {\tt SLHA+} functionality of \CalcHep \cite{Belanger:2010st}, i.e.\ the information about the numerical values of the current parameter space point is passed to \MO via the output file written by \SPheno in the SUSY LesHouches format \cite{Allanach:2008qq}. 

For our scans we have used \SSP which is optimized for parameter scans using the environment provided by model implementations in \SPheno and \MO based on the \SARAH output. \SSP includes also routines to run Markov Chain Monte Carlos (MCMC) in order to find points in parameter space with a very high likelihood according to some defined constraints. 

\section{Exploring the GNMSSM}

In the following we will present the results of our scans over the parameters of the GNMSSM. We will start with the universal case, the CGNMSSM, and subsequently 
relax the requirement of universality. 
We are particularly interested in regions which allow for a rather large Higgs mass. The largest Higgs masses can be achieved when the additional tree-level contribution
to the Higgs mass is large, corresponding to large $\lambda$, (which implies smallish $\kappa$ \cite{Ellwanger:2006rm}) and small $\tan \beta$.
We randomly scan over all the free parameters within this region.

\subsection{The CGNMSSM}
\begin{figure}[!h!]
\centering
\includegraphics[width=0.44\linewidth]{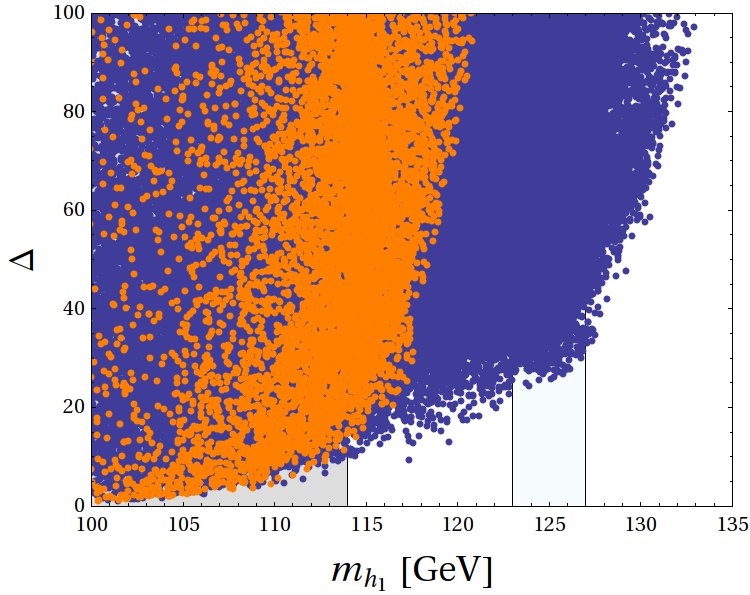}  
\includegraphics[width=0.44\linewidth]{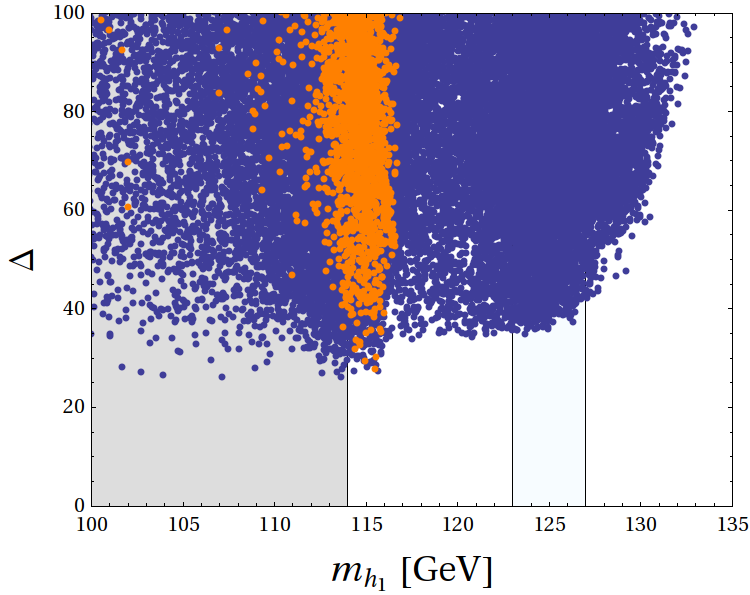}  
\includegraphics[width=0.44\linewidth]{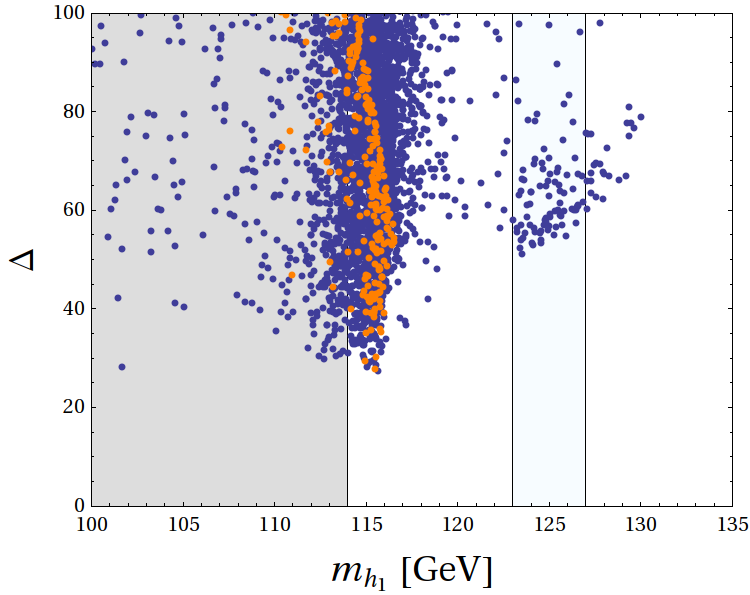} 
\caption{Fine-tuning vs.\ the lightest Higgs mass in the CGNMSSM case with universal boundary conditions 
for the CMSSM (orange), and the CGNMSSM (blue). The first plot corresponds to the unconstrained case, the second plot takes into account the
LHC bounds on particle masses with a cut on squark and gluino masses of $1.2\tev$ and the third plot  assumes an additional upper bound on the
neutralino relic density.}
\label{fig:universal}
\end{figure}
In Figure~\ref{fig:universal} we show the fine tuning as a function of the lightest Higgs mass for both the CMSSM and the CGNMSSM, 
with and without imposing the constraints on the superpartner mass spectrum from the LHC and constraints from the relic density. 
All points have been subjected to internal consistency (e.g.\ no tachyons, correct EWSB etc.).
In the case of the CNMSSM it is known that the requirement of universality is highly restrictive: both $\lambda$ and $\kappa$ have to be rather small and the Higgs mass 
is even smaller than in the MSSM case. We therefore consider the NMSSM case only for more general boundary conditions in the next section. 
Figure~\ref{fig:universal} shows that if no experimental constraints are imposed, the smallest fine tuning is achieved for small Higgs masses, as naively expected.
However, when the parameter space is subjected to the experimental bounds, the lowest fine tuned part of parameter space is not accessible, which is the well 
known ``little hierarchy'' problem.

As may be seen from Figure~\ref{fig:universal} the little hierarchy problem is significantly alleviated in the CGNMSSM for Higgs masses above 118~GeV and a large section of parameter space with only a mild tuning remains.  Interestingly, larger Higgs masses up to $125\;\gev$ do as well in terms
of fine tuning. 
Given the vastness of the CGNMSSM parameter space, we did an iterative scanning procedure, randomly scanning a large part of the parameter space and then
zooming into the interesting looking regions in several steps.
Therefore the points shown in Figure~\ref{fig:universal} are not smoothly distributed over all of parameter space and one should not interpret
the density of points as the probability of finding a viable point with a given fine tuning and Higgs mass.

\begin{figure}[!h!]
 \centerline{
 \includegraphics[width=0.44\linewidth]{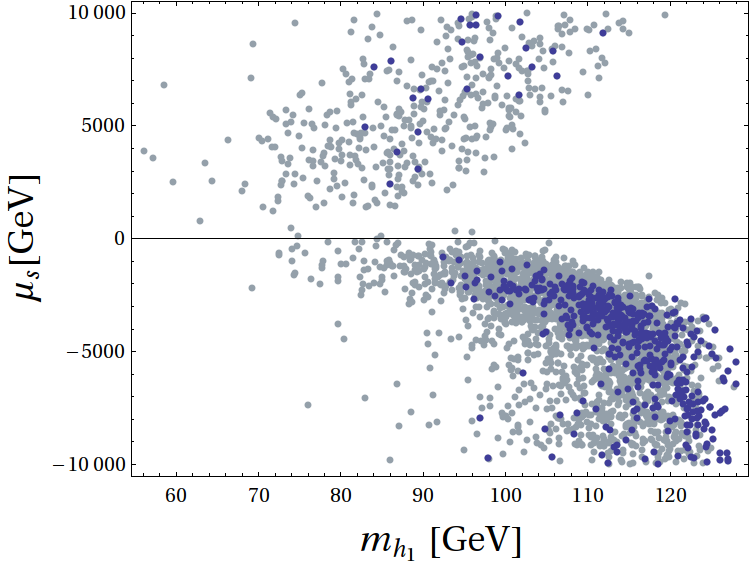}
}
 \caption{The lightest Higgs mass $m_{h_1}$ vs.\ the SUSY conserving singlet mass parameter $\mu_s$ for a representative part
 of parameter space with fine tuning below 100. The light (dark) blue points are before (after) SUSY cuts are taken into account.}
 \label{fig:mus}
 \end{figure}

Of particular interest for the GNMSSM phenomenology is the SUSY conserving singlet mass parameter $\mu_s$, which sets the overall mass scale for the
singlet and the singlino. For the CGNMSSM we show the lightest Higgs mass as a function of this parameter in Figure~\ref{fig:mus}.
It can be seen that for Higgs masses above $120\gev$, the GUT scale parameter $\mu_s$ is rather large, implying that the phenomenology 
of the CGNMSSM is rather `CMSSM like', as also observed in \cite{Ross:2011xv}.
The difference to the CMSSM case is of course a larger Higgs mass and smaller fine tuning.
We show three benchmark points of the CGNMSSM in Table~\ref{tab:benchmark}. 
After SUSY cuts are taken into account, for the Higgs mass region of interest the lowest fine tuned points we find have a fine tuning of about 35.

\subsubsection*{Dark matter}
If we require that the lightest supersymmetric particle (LSP) constitutes the dark matter of our Universe, additional constraints apply. 
In this work we will require a neutralino LSP \footnote{Another interesting scenario would be a gravitino LSP, which could make the region of parameter space
corresponding to a stau NLSP viable.}
and that the relic density does not exceed the 5$\sigma$ WMAP-7 \cite{Jarosik:2010iu} upper bound of $\Omega h^2 \le 0.1298$.
While an under-abundance could always be compensated by the relic density of a multitude of other particles, an overabundance would require a deviation from the standard thermal history of the Universe (or at least a sufficiently
low reheating temperature, such that the dark matter candidate never reaches thermal equilibrium).
It has been shown that it is not possible in the MSSM to get the preferred relic density and a fine tuning below 100 for Higgs masses above 120~GeV \cite{Cassel:2011tg}. 
When we require the correct relic density for the CGNMSSM in addition to the LHC and LEP limits, the lowest fine tuning we find is about 50. 
For a large part of parameter space the lightest neutralino is rather bino-like and dark matter is overproduced, similarly to the CMSSM case.
We find that all points that pass the relic density requirement correspond to the stau coannihilation region.  A region similar to the focus point of the MSSM with a large Higgsino fraction of the LSP doesn't show up. The reason is that the region of interest in the GNMSSM corresponds to small values of $\tan\beta$ while the focus point prefers moderate or large values \cite{Jurcisinova:2005vg}. The direct detection cross section is typically below $5\cdot 10^{-47}\;\text{cm}^{2}$, well below the sensitivity of current dark matter searches.

\subsection{Generalised boundary conditions} 
\label{sec:general}

In this section we will relax the universality condition on the GUT scale parameters and allow the Higgs and singlet soft masses $m^2_{h_d}$, $m^2_{h_u}$ and $m^2_s$ as well as the trilinear parameters $A_\lambda$
and $A_\kappa$ to vary independently. 
\begin{figure}[hbt]
\centering
\includegraphics[width=0.44\linewidth]{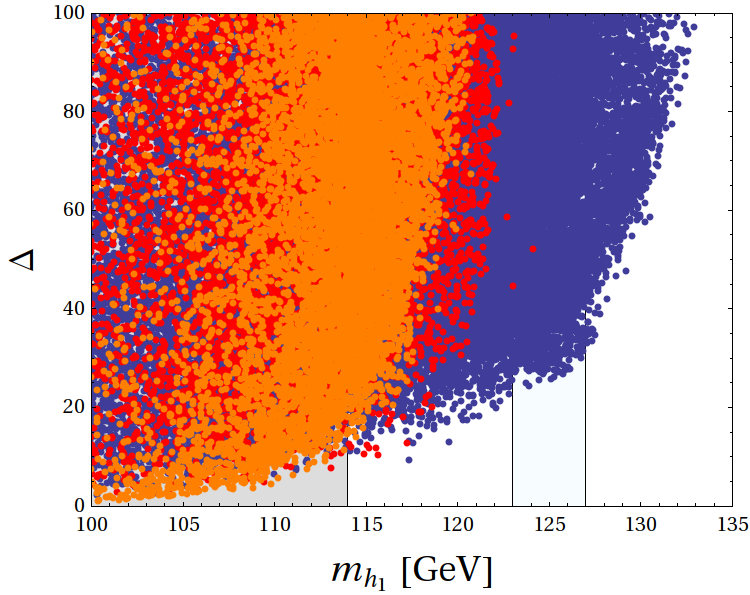}  
\includegraphics[width=0.44\linewidth]{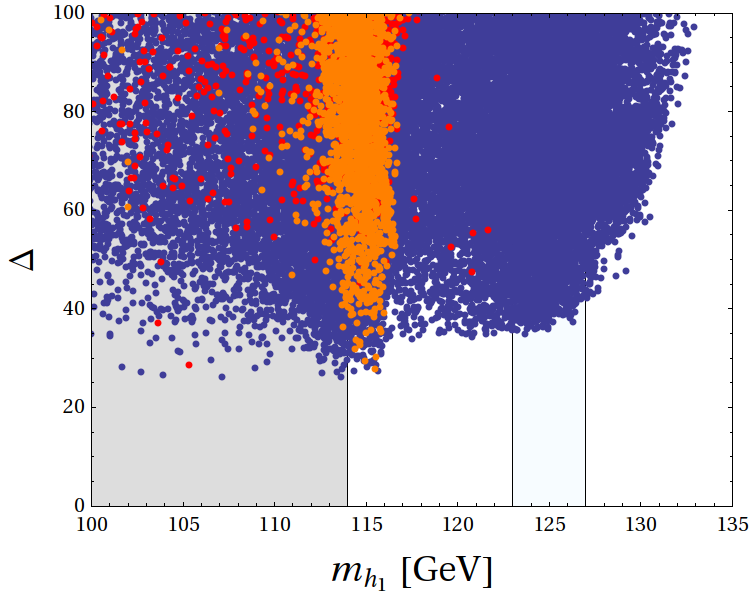}
\includegraphics[width=0.44\linewidth]{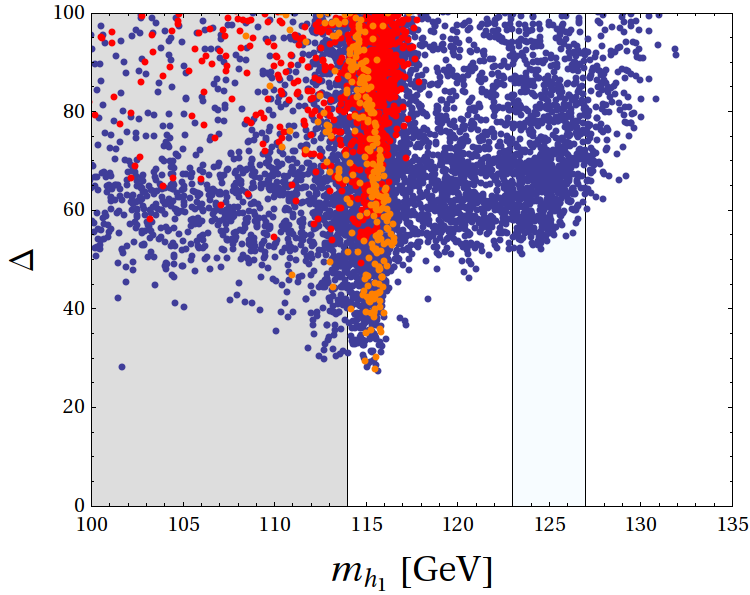}  
\caption{Fine-tuning vs.\ the lightest Higgs mass for the generalised boundary conditions for the MSSM (orange), the NMSSM (red) and the GNMSSM (blue). 
The first plot corresponds to the unconstrained case, the second plot takes into account the
LHC bounds on particle masses and the third plot assumes an additional upper bound on the
neutralino relic density.}
\label{fig:nonuniversal}
\end{figure}
In Figure~\ref{fig:nonuniversal} we show the fine tuning as a function of the lightest Higgs mass, this time for the 
GNMSSM as well as the MSSM and NMSSM.
For the GNMSSM we find no significant improvement of the fine tuning with respect to the universal case.
Nevertheless, for large Higgs masses, the fine tuning is still significantly lower than in the MSSM or NMSSM case.
One interesting difference compared to the universal case is the freedom to have small (in the NMSSM case zero) values for the singlet mass
parameter $\mu_s$. Therefore the singlet states can be light, leading to potentially interesting phenomenology.
One interesting possibility is that the lightest Higgs is mainly singlet and the second lightest Higgs corresponds to the
mainly MSSM-like state. 
\begin{figure}[hbt]
\centerline{
\includegraphics[height=6cm]{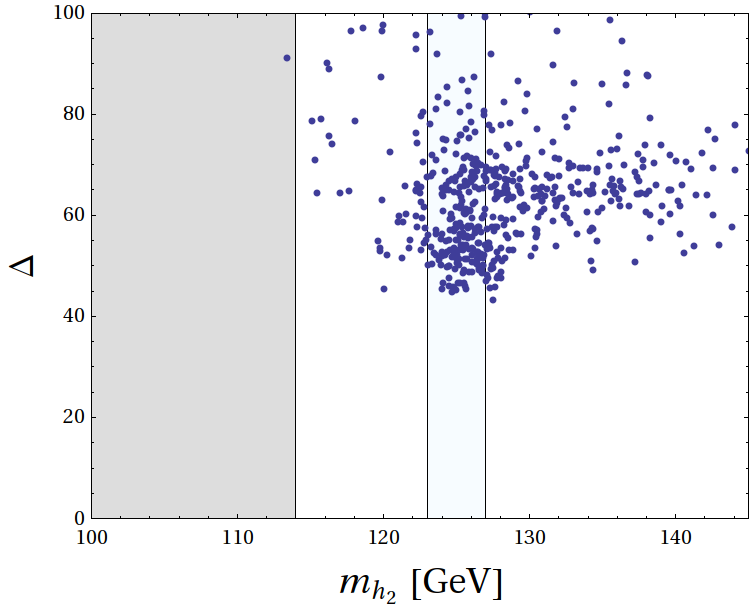}}
\caption{The second lightest Higgs  mass vs.\ the fine tuning in the case the lightest Higgs is mostly singlet and evades all experimental bounds.
The imposed cuts are $1.2\tev$ on squark and gluino masses and an upper bound on the relic density according to WMAP.}
\label{fig:singlet}
\end{figure}
In Figure~\ref{fig:singlet} we show the second lightest Higgs mass vs.\ the fine tuning for the case the lightest Higgs is mostly 
singlet and evades all experimental bounds.
We find that a light singlet can be an advantage for the dark matter abundance in that it seems much easier to reduce the relic abundance. In our scans much more points showed up with low fine-tuning, the correct
relic density and with Higgs and squark masses in the preferred ranges for the case
where the lightest scalar has a large singlet component. The origin
of this is most likely the much weaker correlation between the singlino mass and annihilation
cross section and the other masses and electroweak parameters.
Correspondingly we are no longer restricted to the narrow stau coannihilation region.
Another interesting observation is that the direct detection cross-section is close to current bounds and hence this part of GNMSSM
parameter space will be probed not only by the LHC but also by the next generation of direct detection experiments. 
We show two benchmark points for the generalised boundary conditions in Table~\ref{tab:benchmark}.
\begin{table}[htb]
\centering
 \begin{tabular}{|c|c c c c c|}
\hline
 & BP1 & BP2 & BP3 & BP4 & BP5\\
\hline
$m_0$ [GeV]                                 &746                & 163                        & 957                     & 573                    &752\\
$m_{1/2}$ [GeV]                             &476                & 568                        & 557                     & 482                    &472\\
$\tan \beta$                                &2.7                &2.9                         & 2.8                     & 3.4                    &2.8\\
$A_0$ [GeV]                                 &1433               &1666                        & 782                     & 27                     &-198\\
$\lambda$                                   &1.43               &1.47                        & 1.58                    & 1.34                   &1.12\\
$\kappa$                                    &-0.1               &0.09                        & -0.005                  & 1.52                   &1.03\\
$A_\lambda$ [GeV]                           &$A_0$              & $A_0$                      & $A_0$                   & 400                    &192\\
$A_\kappa$  [GeV]                           &$A_0$              & $A_0$                      & $A_0$                   & -323                   &-326\\      
$v_s$       [GeV]                           &-841               & -190                       & -929                    & 390                    &281\\
$\mu_s$     [GeV]                           &-5931              &-5354                       & -5799                   & 131                    &-37\\
$m_{h_d}^2~[\text{GeV}^2]$                  & $m_0^2$           & $m_0^2$                    &  $m_0^2$                &  $9.1  \cdot 10^{5}$   &$5.4\cdot 10^{5}$\\
$m_{h_u}^2~[\text{GeV}^2]$                  & $m_0^2$           & $m_0^2$                    &  $m_0^2$                &  $2.3  \cdot 10^{6}$   &$2.4  \cdot 10^{6}$\\
$m_s^2~[\text{GeV}^2]$                      & $m_0^2$           & $m_0^2$                    &  $m_0^2$                & $2.8 \cdot 10^{6}$     & $1.7\cdot 10^{6}$\\
$\mu$  [GeV]                                & -750              & -1136                      &  -934                   & -33                    &10\\
$b\mu~[\text{GeV}^2]$                       & $-2.4\cdot 10^6$  & $-1.2\cdot 10^6$           &  $-2.3\cdot 10^6$       & 147                    &26\\
$b_s~[\text{GeV}^2]$                        & $-1.9\cdot 10^7$  & $-5.4\cdot 10^6$           &  $-1.4\cdot 10^7$       & 326                    &144\\
$\xi_s~[\text{GeV}^3]$                      & $2.2\cdot 10^9$   & $1.5\cdot 10^9$            &  $3.0\cdot 10^9$        & 22                     &-8\\
\hline  
$m_\text{squark}$ [GeV]                     &1256-1293               &1207-1263              &  1507-1548              &1211-1248               &1280-1315\\
$m_{\tilde{g}}$  [GeV]                      &1219                    &1389                   &  1416                   &1242                    &1235 \\
\hline                         
$m_{h_1}$ [GeV]                             &124                     &123.5                  & 125                     & 93.5                   &78\\
$m_{h_2}$ [GeV]                             &1002                    &856                    & 1257                    & 125                    &124\\
$h_1$ singletfraction                       & $\mathcal{O}(10^{-4})$ &$\mathcal{O}(10^{-6})$ & $\mathcal{O}(10^{-4})$  & 0.8                    &0.85\\
$\text{Br}(h \rightarrow \gamma \gamma)$    & $2.29 \cdot 10^{-3}$   &  $2.28 \cdot 10^{-3}$ &    $2.2 \cdot 10^{-3}$  & $2.5 \cdot 10^{-3}$    &$2.66 \cdot 10^{-3}$\\
$\text{Br}(b \rightarrow s \gamma)$         & $3.1 \cdot 10^{-4}$    & $3.1 \cdot 10^{-4}$   & $3.1 \cdot 10^{-4}$     & $3.1 \cdot 10^{-4}$    &$3.3 \cdot 10^{-4}$\\
$\Delta a_\mu$                              & $-7.8 \cdot 10^{-11}$  & $-2.5 \cdot 10^{-10}$ & $-5.4 \cdot 10^{-11}$   & $1.7 \cdot 10^{-10}$   &$8 \cdot 10^{-11}$ \\
$\delta\rho$                                &  $6.2 \cdot 10^{-5}$   &   $6.6 \cdot 10^{-5}$ &  $7.5 \cdot 10^{-5}$    &  $1.9 \cdot 10^{-4}$   &$3.1 \cdot 10^{-4}$  \\
\hline
$m_{\tilde{\chi}^0_1}$ [GeV]                &229                     &270                    & 168                     & 99                     &70\\
$\tilde{\chi}^0_1$ singlinofraction         & $\mathcal{O}(10^{-5})$ & $\mathcal{O}(10^{-5})$&  $\mathcal{O}(10^{-5})$ & 0.1                    &0.2\\
$\Omega h^2$                                &  7.5                   &0.10                   & 7.4                     & 0.017                  &0.11\\
$\sigma_p [cm^2]$                           & $2.8 \cdot 10^{-47}$   & $2.2 \cdot 10^{-47}$  &  $6 \cdot 10^{-47}$     &  $1.2 \cdot 10^{-44}$  & $1.3 \cdot 10^{-45}$\\
\hline
$\Delta$ (Fine-tuning)                      &34.9                    &51.0                   & 51.8                    & 44.9                   &52.7\\
\hline
 \end{tabular}
\caption{Benchmark scenarios for the GNMSSM for the universal (BP1-BP3) and the general (BP4-BP5) case. $m_\text{squark}$ shows the range of squark masses
         of the first two generations. For the last two points the second lightest Higgs is mostly MSSM-like. All input parameters except $\tan\beta$ and $v_s$ are given at the GUT scale.}
\label{tab:benchmark}
\end{table}

\subsection{Comparison between the NMSSM and the GNMSSM}
\begin{figure}[!h!]
\centering
\includegraphics[width=0.44\linewidth]{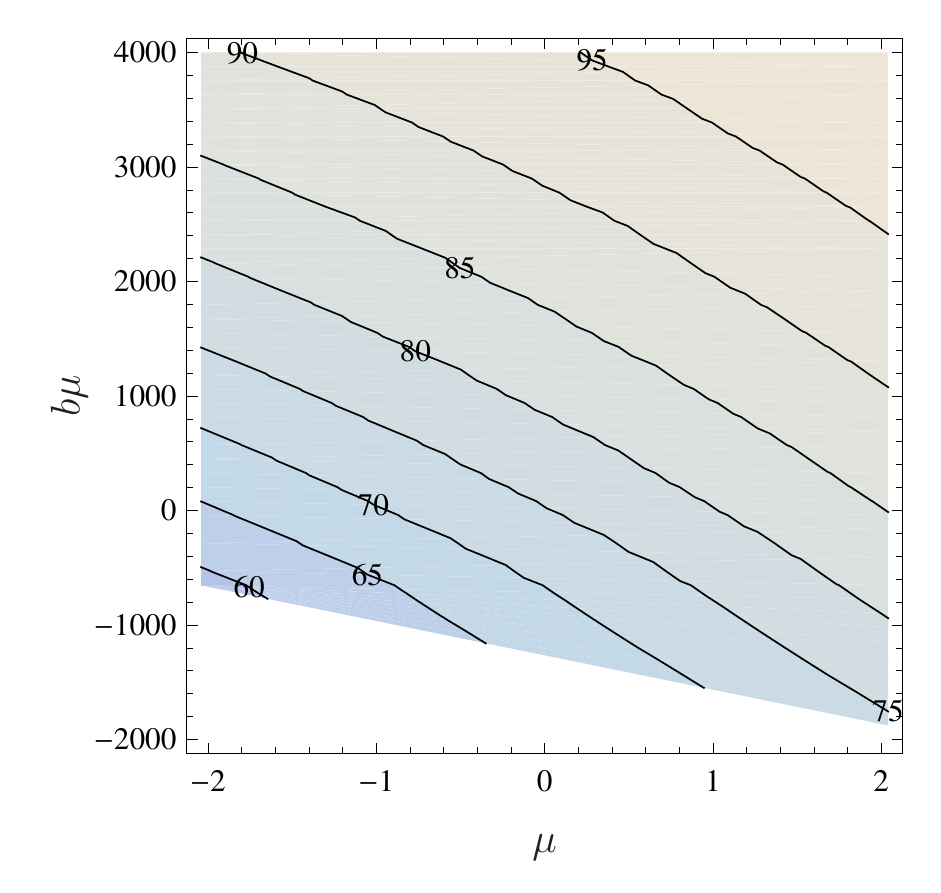}  
\includegraphics[width=0.44\linewidth]{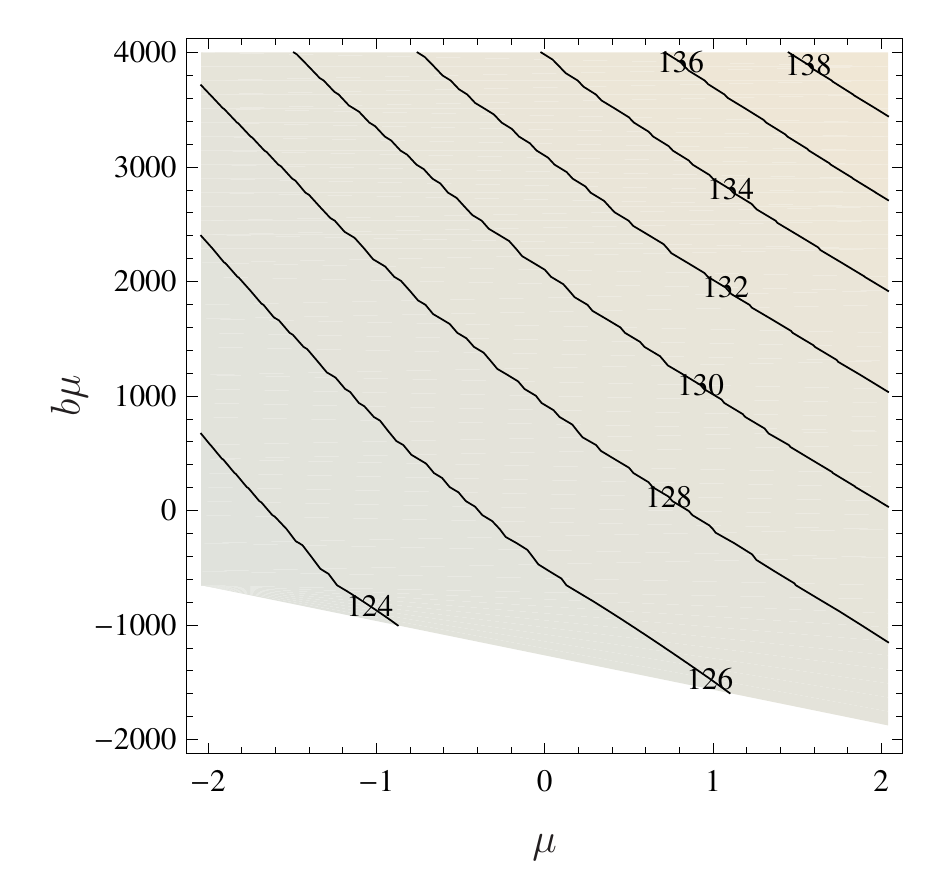} \\
\includegraphics[width=0.44\linewidth]{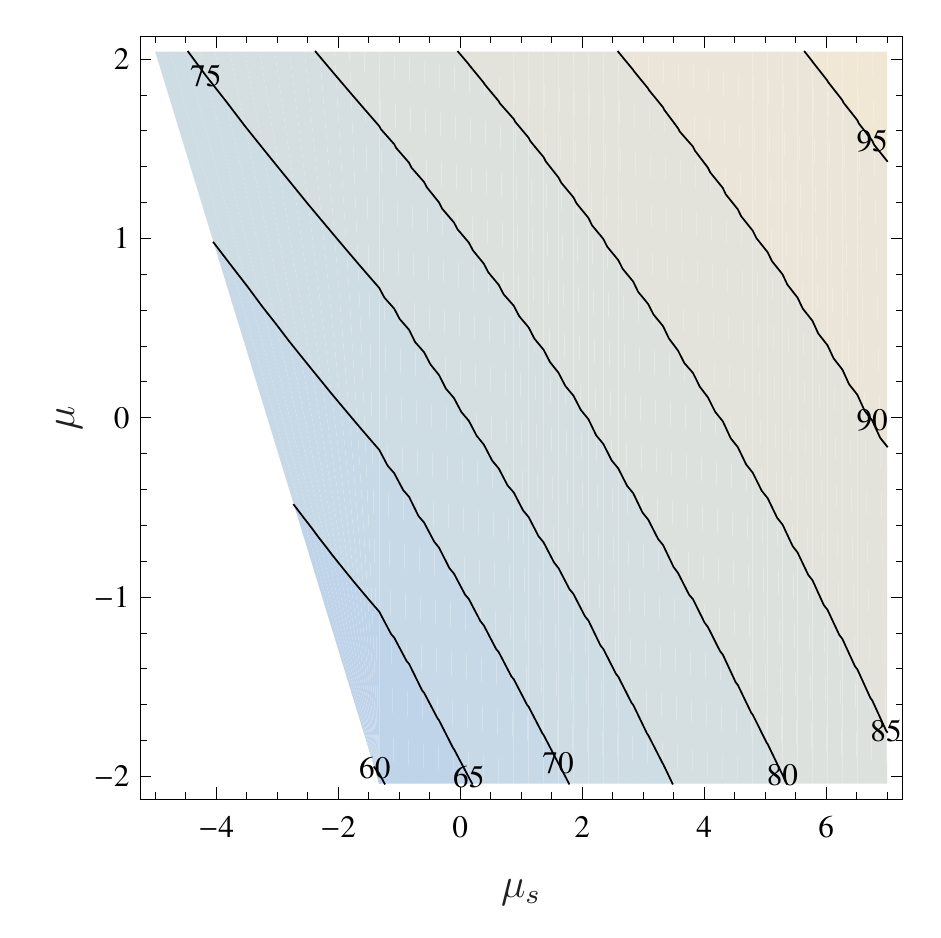}
\includegraphics[width=0.44\linewidth]{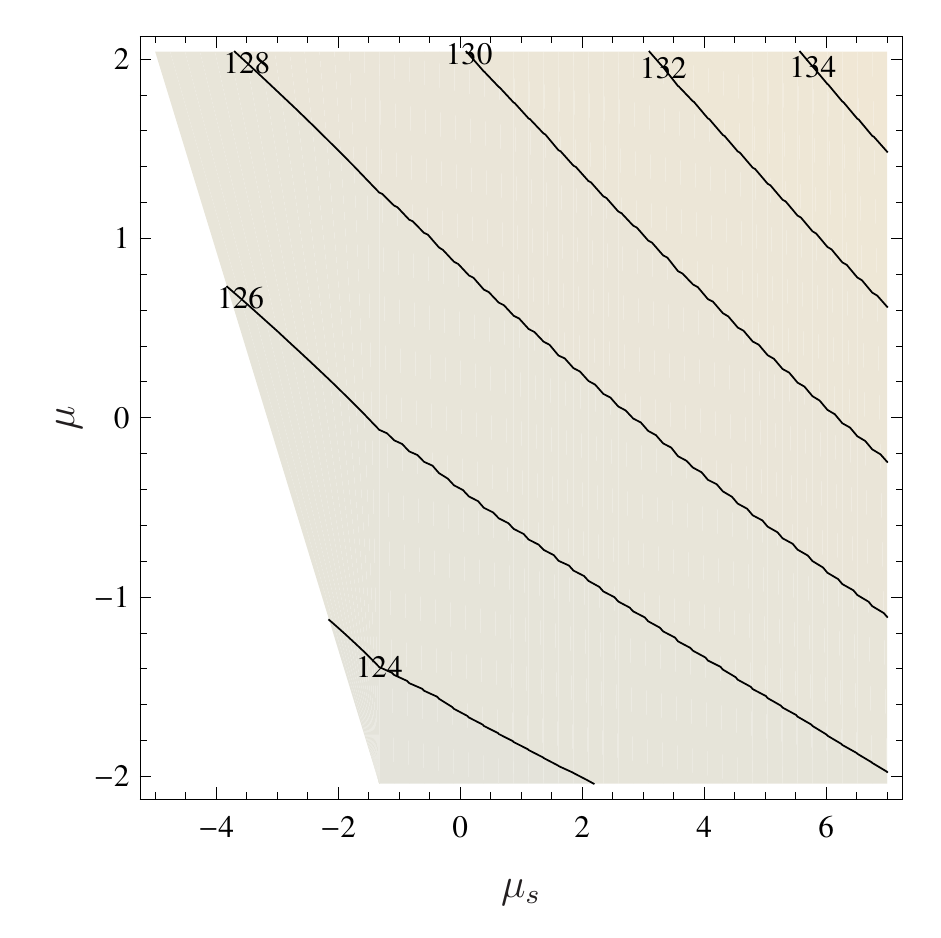}    
\caption{The mass of the lightest, singlet-like Higgs (left) and second lightest, MSSM-like Higgs (right) for a variation of the GNMSSM specific parameters $(\mu, b_\mu)$ (first row) and $(\mu_s, \mu)$ (second row). The other parameters have been chosen as in~\cite{Ellwanger:2012ke}: $m_0 = 385$~GeV, $m_{1/2} = 430$~GeV, $\tan\beta = 3.0$, $A_0 = 1590$~GeV, $\lambda=1.23$, $\kappa = 1.07$, $A_\lambda = 1580$~GeV, $A_\kappa = 1560$~GeV and $v_s = 250$~GeV and the GNMSSM parameters not shown in the different Figures were kept~0.}
\label{fig:NMSSMvsGNMSM}
\end{figure}

It is interesting to explore the effect of the additional independent parameters of the GNMSSM compared to the NMSSM for the Higgs masses.  To do this we have chosen to analyse a NMSSM point discussed in Ref.~\cite{Ellwanger:2012ke} with a Higgs mass near 126~GeV and  perturb away from the NMSSM point via the additional GNMSSM parameters. The results for a variation of $(\mu,b \mu)$ and $(\mu_s,\mu)$ are depicted in Figure~\ref{fig:NMSSMvsGNMSM}. One may see that even small values of the new parameters can change the Higgs masses significantly. Since the largest contributions of the fine tuning are due to the CMSSM parameters $m_0$, $m_{1/2}$ and $A_0$, the variation of the parameters in the Higgs sector has only a very small impact on the overall fine tuning. Therefore, it is possible to vary the Higgs masses using these parameter without increasing the fine tuning.  

\section{Phenomenological aspects of the GNMSSM}

The extension of the MSSM to include a singlet state can significantly change the phenomenology. The most important differences come about due to the possibility that there is an additional light Higgs state that is mainly singlet and the possibility that the LSP is mainly composed of the singlino.
For the case of universal scalar boundary conditions the lowest fine tuned points correspond to large $\mu_{s}$ (c.f.\ Figure \ref{fig:mus}) and in this case the singlet states are heavy and the phenomenology is close to MSSM phenomenology but with a Higgs that can be much heavier. However, for the case of generalised boundary conditions discussed in Section~\ref{sec:general}, low-fine tuned points can have small $\mu_{s}$  and hence different phenomenology while still having a heavy Higgs state consistent with the LHC hints for a $125$~GeV Higgs.

\subsection{Higgs phenomenology}
\begin{figure}[!h!]
\centering
\includegraphics[width=0.44\linewidth]{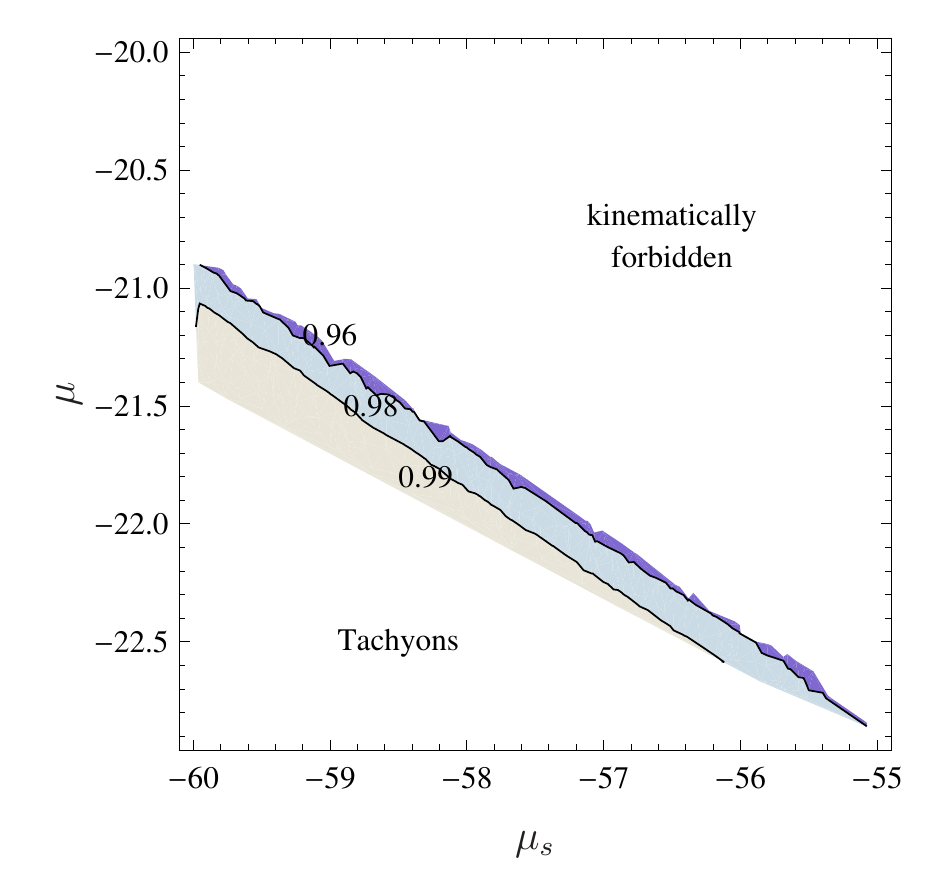}  
\includegraphics[width=0.44\linewidth]{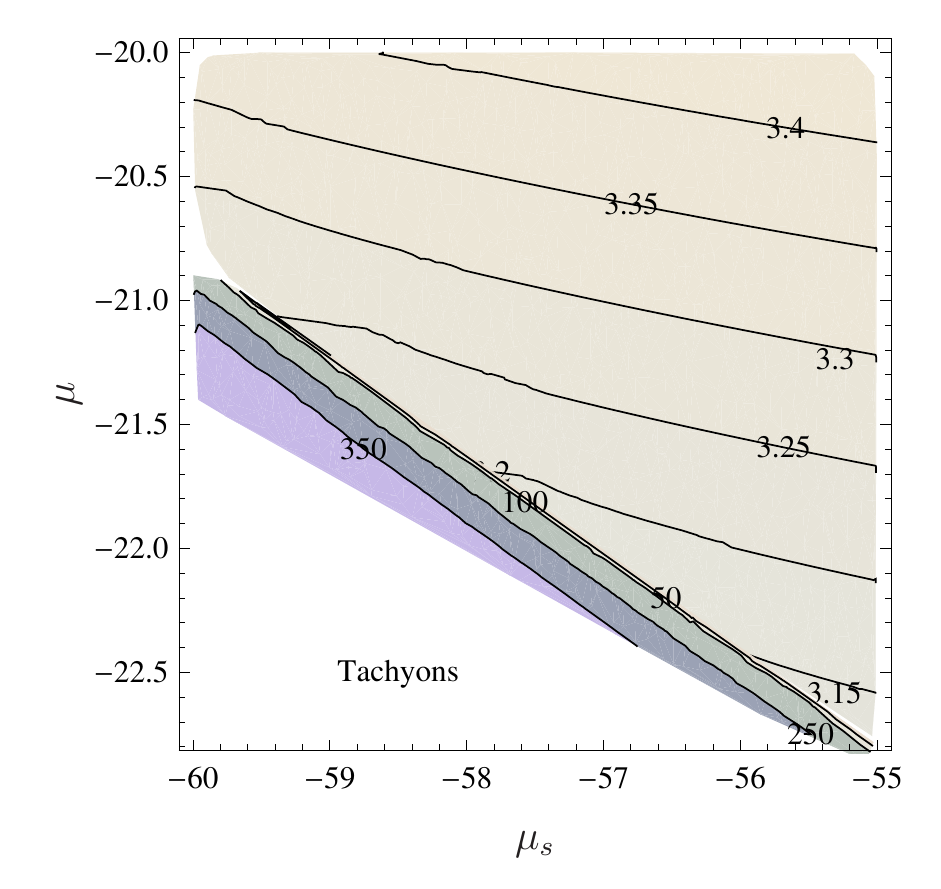}
\caption{Decay properties of the MSSM-like Higgs for a variation of $\mu$ and $\mu_s$. Left: $\text{Br}(h_2 \to h_1 h_1)$ (left), right: total width of $h_2$ (right) in MeV. The other parameters were set to $m_0 = 640$~GeV, $m_{1/2} = 480$~GeV, $\tan\beta = 2.53$, $A_0 = -253$~GeV, $\lambda=0.98$, $\kappa = 1.19$, $A_\lambda = 136.2$~GeV, $A_\kappa = -410.3$~GeV, $v_s = 620.8$~GeV, $b \mu = 95.2$~GeV${}^2$, $b_s = 215.2$~GeV${}^2$ and $\xi_s = -106.5~\text{GeV}^3$. The overall fine tuning for the shown parameter range is $\sim$ 58 - 60.}
\label{fig:H2toH1H1}
\end{figure}

\begin{figure}[!h!]
\centering
\includegraphics[width=0.32\linewidth]{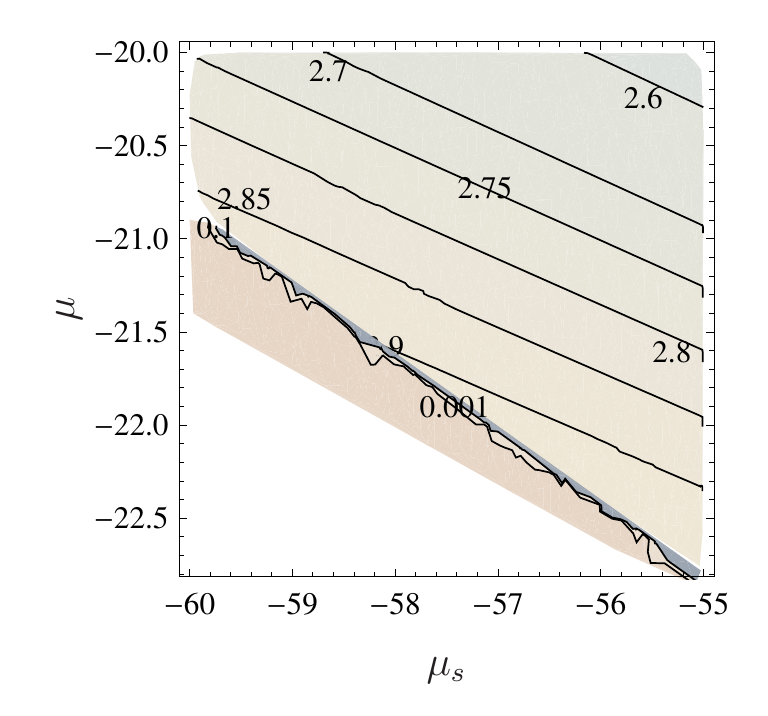}
\includegraphics[width=0.32\linewidth]{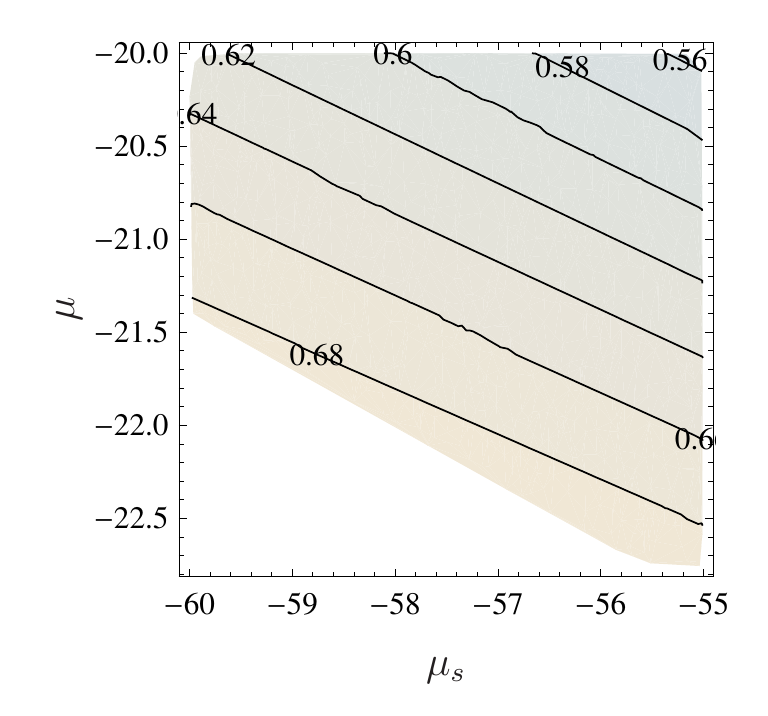}
\includegraphics[width=0.32\linewidth]{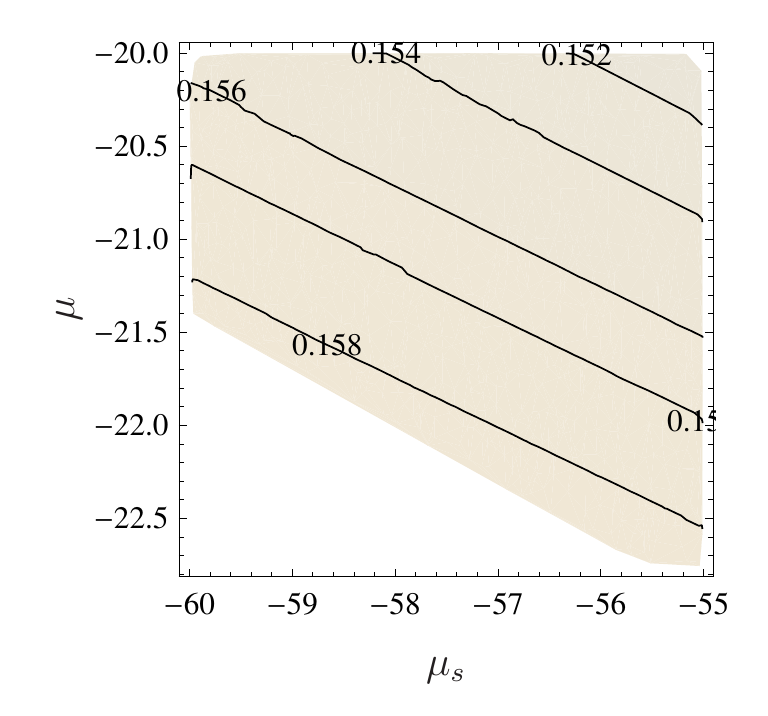}        
\caption{Photonic decay of the MSSM-like Higgs. The left plot shows $\text{Br}(h_2 \to \gamma \gamma) \times 1000$. The other two figures show the up-type (middle) and down-type (right) fraction of the light MSSM-like Higgs. The parameters are the same as for Figure~\ref{fig:H2toH1H1}. Comparing to Figure~\ref{fig:H2toH1H1} we see that the region where
the photonic decay becomes very suppressed corresponds exactly to the region where the decay channel $h_2 \to h_1 h_1$ opens up.}
\label{fig:H2toPP}
\end{figure}

Let us first consider the effect of a light, mainly singlet, Higgs, $h_{1}$, on Higgs phenomenology. It can change the decay properties of the MSSM-like Higgs field dramatically as is depicted in Figure~\ref{fig:H2toH1H1}. Depending on the values of $\mu$ and $\mu_s$ the decay of the mainly doublet Higgs, $h_{2}$, into two singlet Higgs fields can be kinematically allowed. If this is the case the branching ratio $\text{Br}(h_2 \to h_1 h_1)$ will be very large and the total width of the MSSM Higgs will typically be increased by more than one order of magnitude. 
Another interesting feature is that the up-type fraction of the MSSM-like Higgs is particularly sensitive to the $\mu$ parameter without significantly changing the down-type fraction. This allows for an increase $\text{Br}(h_2 \to \gamma \gamma)$ as may be indicated by the LHC measurements. As shown in the left plot of Figures~\ref{fig:H2toPP}, changing $\mu$ from -20.9~GeV to -22.5~GeV (for $\mu_S = -55$~GeV) enhances the branching ratio of the di-photon channel from less than $2.6\cdot 10^{-3}$ to more than $2.9\cdot 10^{-3}$. The reason is that the up-type fraction of the doublet Higgs changes by more than 10\%.

\begin{figure}[!h!]
 \centerline{  \includegraphics[width=0.44\linewidth]{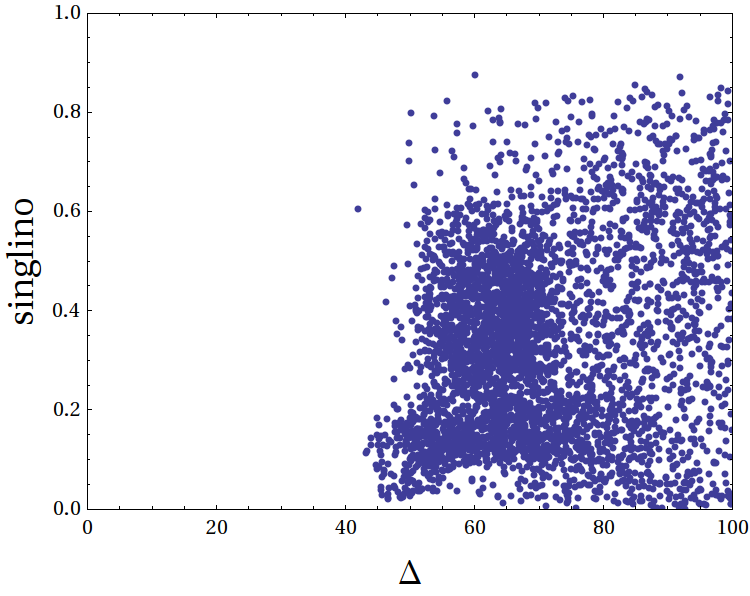}}
 \caption{Singlino fraction of the LSP vs.\ fine tuning}
 \label{fig:FT_singlino}
 \end{figure}
 
 \begin{figure}[!h!]
\centerline{
\includegraphics[height=0.3\linewidth]{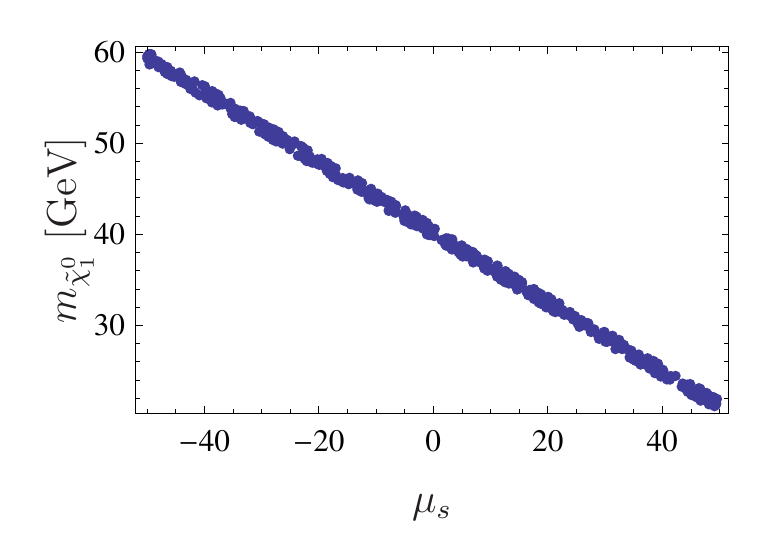}
\includegraphics[height=0.295\linewidth]{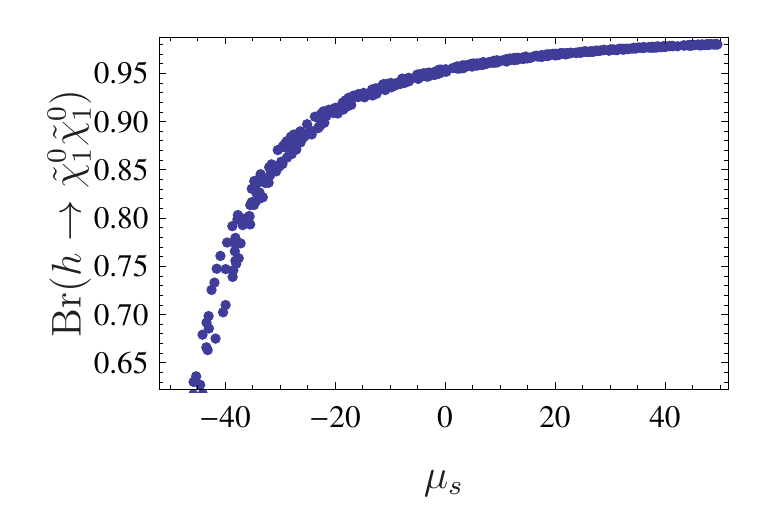}}
 \caption{Decay of the light Higgs particle into two singlinos as a function of $\mu_s$. The left plot shows the mass of the LSP as function of $\mu_s$, whereas the right plot gives the branching ratio Br($h \to \tilde{\chi}^0_1 \tilde{\chi}^0_1$). The other parameters have been chosen as $m_0 = 525$~GeV, $m_{1/2} = 591$~GeV, $\tan\beta = 2.9$, $A_0 = 1410$~GeV, $\lambda=1.42$, $\kappa = [0.04.,0.05]$, $A_\lambda = -1815$~GeV, $A_\kappa = -569$~GeV, $v_s = -360$~GeV, $\mu = 60$~GeV, $b \mu = 4056$~GeV${}^2$, $b_s = -6545$~GeV${}^2$ and $\xi_s = -144174~\gev^3$. }
 \label{fig:H_to_N1N1}
 \end{figure}

Higgs phenomenology can also be affected by the change allowed in the GNMSSM in the neutralino structure, particularly in the nature of the LSP. As may be seen in  Figure~\ref{fig:FT_singlino} there is minimal correlation between the singlino fraction or the singlino mass and the fine tuning. As a result it is possible to find areas in parameter space with low fine tuning and $m_\text{LSP} < \frac{1}{2} m_h$. 
In this case the invisible decay $h \to \tilde{\chi}^0_1 \tilde{\chi}^0_1$ is possible. Again, as shown in Figure~\ref{fig:H_to_N1N1}, the exact branching ratio is also very sensitive to the parameters, particularly the value of $\mu_s$. A variation of $\mu_s$ in the range [-50,50]~GeV changes the mass of the MSSM-like Higgs only by 0.5~GeV, but can increase the branching ratio into two singlinos from less than 60\% to nearly 100\%. This is mainly caused by the increase in phase space due to the decrease in the mass of the singlino. 
If the indications of a 125 GeV Higgs turned out to be true, these Higgs decays to lighter singlet states would be strongly constrained and the likely case that they are kinematically forbidden
would be preferred. On the other hand, if the hints turn out to be a statistical fluctuation, these decays would offer interesting escape routes for Higgs physics.
A more detailed study of these and other interesting aspects of the Higgs sector in the GNMSSM will be given elsewhere \cite{SchmidtHoberg:2012yy}.

\subsection{SUSY phenomenology}
The nature of supersymmetric signals is largely determined by the LSP. For the case that $\mu_{s}$ is large the LSP is as in the MSSM so the signatures will be the same as in the MSSM. For the case that $\mu_{s}$ is small and the LSP is mainly  singlino the signatures can change. 

The LHC is sensitive to gluino and squark pair production. Let us consider first the case the gluino is lighter than the squarks. 
The dominant gluino decay mode is likely to be the three body decay into a quark anti-quark pair and the LSP. The singlino has no direct coupling to squarks so this decay will proceed via its neutral gaugino or Higgsino component. Over a large part of parameter space the singlino dominated LSP has a significant Higgsino component so the dominant decay mode will be into the third generation of quarks, $\tilde{g}\rightarrow t\bar{t}\tilde{s}, b\bar{b}\tilde{s}$. Note that this expectation is different from the MSSM because the dark matter constraints on the makeup of the LSP are different. In the MSSM the LSP cannot be dominantly Higgsino and the gaugino component (mainly Bino) dominates. In the GNMSSM the singlino dominates but the Higgsino component is typically much larger than the gaugino component. 
Thus the collider signal for gluino pair production will be four jets associated with the third generation of quarks plus missing energy and momentum.
The dominant decay of squarks is first to a quark and the gluino followed by the gluino decay into third generation quarks. Thus the signal for squark pair production will be multijets mostly associated with third generation quarks plus missing energy and momentum.

For the case the squarks are lighter than the gluino their dominant decay mode will be the two body decay into the associated quark and singlino. Thus the signal for squark pair production in this case will be di-jet production plus missing energy and momentum.
Gluinos will dominantly decay to a quark (anti)squark pair giving a final state of quark antiquark plus LSP. Thus the signal for gluino pair production in this case are fourquark-jets plus missing energy and momentum. 

\section{Summary and Conclusions}
The realisation that simple discrete $R$ symmetries naturally constrain the $\mu$ and $\mu_{s}$ parameters to be of the SUSY breaking scale mean that the GNMSSM is as natural a theory as the NMSSM.  Indeed the fact that the symmetry also eliminates the dangerous dimension three, four {\it and} five baryon- and lepton-number violating terms in the Lagrangian and avoids destabilising tadpoles and domain  wall problems renders it a more promising starting point than the NMSSM. 

In this paper we have shown that the GNMSSM, with gravity mediated SUSY breaking and universal boundary conditions, can accommodate a Higgs mass up to $130$~GeV without a significant increase in the lowest fine tuning needed which remains about 1 part in 35. This is in contrast to the MSSM which requires a fine tuning of greater than 1 part in 300  to accommodate such a Higgs mass. 
In this work we have assumed an underlying GUT structure at a scale around $10^{16} \gev$. In particular we have assumed universal squark masses $m_0$ at the high scale.
If we were to relax this assumption, even smaller values of the fine tuning should be achievable, both due to a potentially lighter stop as well as due to a lower scale 
at which the fine tuning is evaluated, cf.~e.g.~\cite{Hall:2011aa}.
We have also determined the dark matter abundance and shown that a significant proportion of the low fine tuned parameter space can lead to dark matter close to the WMAP bound without a significant increase in the fine tuning. The dark matter lies in the stau co-annihilation region and its direct detection cross section is far below the sensitivity of direct detection dark matter experiments. The low fine tuned points correspond to large $\mu_{s}$. In this region the new singlet states are heavy and the phenomenology is very close to that of the MSSM but with a heavier Higgs allowed.

In order to make a comparison with the NMSSM we have also explored the more general case relaxing the universality condition on the GUT scale parameters and allowing the Higgs and singlet soft masses $m^2_{h_d}$, $m^2_{h_u}$ and $m^2_s$ as well as the trilinear parameters $A_\lambda$
and $A_\kappa$ to vary independently. We found the fine tuning in the GNMSSM is relatively insensitive to this change and that it is significantly lower than that found for the NMSSM, particularly in the $125$~GeV Higgs mass range. Interestingly for these more general boundary conditions the fine tuning is relatively insensitive to a reduction in the singlet $\mu_{s}$ term and as a consequence there may be an additional light, mainly singlet, Higgs scalar and the LSP can have a large singlet component. This can change the phenomenology dramatically. The mainly singlet Higgs state can be the lightest state without conflicting with LEP and LHC bounds and indeed the doublet Higgs state can dominantly decay into pairs of singlet Higgs. Alternatively the doublet Higgs state can dominantly decay invisibly to a pair of mainly singlino LSPs. The case that the LSP is mainly singlino also affects the dominant decay channels of the gluino and squarks giving characteristic signals that will provide tests of the scheme. Also a light singlino LSP changes the dark matter expectation as it is much easier to reduce the relic abundance so that we are no longer restricted to the narrow stau coannihilation region. In this case the direct detection cross-section can be close to current bounds and hence this part of GNMSSM parameter space will be probed not only by the LHC but also by the next generation of direct detection experiments.

\section*{Acknowledgements}
We would like to thank Lawrence Hall for useful discussions as well as Werner Porod and Tim Stefaniak for their support concerning {\tt SPheno} and {\tt HiggsBounds}. The research presented here was partially supported by the EU ITN grant UNILHC 237920 (Unification in the LHC era) and the ERC Advanced Grant BSMOXFORD 228169.

\bibliography{GNMSSM}
\bibliographystyle{ArXiv}

\end{document}